\documentclass[
               showpacs,            
               preprintnumbers,     
               aps,                 
               prd,                 
               a4paper,             
               superscriptaddress,  
               unsortedaddress,     
               footinbib,           
               tightenlines,        
               floats               
               ]{revtex4}

\usepackage{graphicx}
\usepackage{latexsym}
\usepackage{amsmath,amssymb}        
\usepackage[draft=false]{hyperref}

\newcommand{\be}{\begin{equation}}
\newcommand{\ee}{\end{equation}}
\newcommand{\bea}{\begin{eqnarray}}
\newcommand{\eea}{\end{eqnarray}}
\def\R{{\cal R}}
\def\s{{\rm s}}

\begin{document}
\title{Cosmological matching conditions}
\author{Edmund J.~Copeland\footnote{email: Ed.Copeland@nottingham.ac.uk}}
\affiliation{School of Physics and Astronomy, University of
  Nottingham, University Park, Nottingham, NG7 2RD}
\author{David Wands\footnote{email: david.wands@port.ac.uk}}
\affiliation{Institute of Cosmology and Gravitation, University of
  Portsmouth, Portsmouth, PO1 2EG} 

\date\today

\pacs{98.80.Cq \hfill hep-th/0609183}

\begin{abstract}
We investigate the evolution of scalar metric perturbations across a
sudden cosmological transition, allowing for an inhomogeneous surface
stress at the transition leading to a discontinuity in the local
expansion rate, such as might be expected in a big crunch/big bang
event.  We assume that the transition occurs when some function of
local matter variables reaches a critical value, and that the surface
stress is also a function of local matter variables.  In particular we
consider the case of a single scalar field and show that a necessary
condition for the surface stress tensor to be perturbed at the
transition is the presence of a non-zero intrinsic entropy
perturbation of the scalar field.  We present the matching conditions
in terms of gauge-invariant variables assuming a sudden transition to
a fluid-dominated universe with barotropic equation of state.  For
adiabatic perturbations the comoving curvature perturbation is
continuous at the transition, while the Newtonian potential may be
discontinuous if there is a discontinuity in the background Hubble
expansion.
\end{abstract}

\maketitle

\section{Introduction}

The evolution of cosmological perturbations through a sudden
transition in the cosmological evolution has been a subject of much
interest over the past decade or so. Starting with the original idea
due to Gasperini and 
Veneziano of the Pre Big Bang scenario, which is based on the
effective four dimensional low energy string action (for reviews see
\cite{GVreview,LWC}), through to the more recent ideas due to
Steinhardt, Turok and collaborators, which is motivated by brane
collisions in M-theory, whether it be through a single collision
arising in the Ekpyrotic scenario \cite{ekpyrotic} -
\cite{Khoury:2001zk}, or as a result of a series of brane collisions
as in the Cyclic scenario \cite{Steinhardt:2001st} -
\cite{Turok:2004gb}, dealing with the propagation of cosmological
perturbations through the collision region has generated considerable
debate.

Of particular concern has been the question of whether the approximately
scale-invariant spectrum of fluctuations as observed in the microwave
background anisotropies \cite{Spergel:2006hy} can be generated in
these type of collapse/re-expansion models, as opposed to the case of
the Inflationary Universe where its origin is generally 
accepted. Addressing these issues is not so straightforward. The
equations of motion can become singular as the transition region is
reached, or can become ill defined in that extra terms may be induced
which we do not have control over. Moreover, most attempts to model
the behaviour of the propagating modes to date have been based on an
effective four-dimensional description of the Universe. In the case of
models relying explicitly on the evolution of branes in an extra fifth
dimension, such an explanation may have its limitations. Indeed in
\cite{Tolley:2003nx,McFadden:2005mq} the authors argue that the extra
dimension plays a vital role in determining the evolution of the
perturbations through the transition. This could be important, there
have been many results published to date, and depending on the
approach taken, whether it be an effective four-dimensional or
five-dimensional transition, there are claims that the spectrum of 
perturbations is either consistent with observations or completely
inconsistent.
In addressing this thorny issue we have to be clear about the
assumptions that are being made, because it could well be that the
apparent inconsistencies in the published results have arisen because
the authors are not in fact discussing the same physical problem.

One of the problems we face in trying to determine the true evolution
of the perturbations is to describe the matching conditions across the
transition. In this short paper we propose an improved method for
achieving this. The system we will be investigating is four
dimensional and although we will be primarily interested in cases
where there is an abrupt change in the expansion rate at the matching
surface, motivated by attempts to match cosmological perturbations
through a ``bounce'' from contraction to expansion, our results may
also be applied to other cosmological transitions such as reheating at
the end of inflation where this may also be modelled by a sudden transition.
We generalise earlier studies of cosmological matching conditions
\cite{HV,DM,MS} by allowing a discontinuity in the expansion due to a
surface stress tensor at the transition,
as suggested by Durrer and Vernizzi \cite{DV}. 
We will require continuity of the metric, but not its normal derivatives.

The method we follow involves introducing a locally defined scalar
function $\Sigma (\chi_I)$, where the moduli $\chi$ represent any
local scalar observable.  At the transition, where the matching is
described across a spatial hypersurface, $\Sigma$ reaches a critical
value. By expanding about this critical value in terms of the moduli,
we derive a set of conditions that have to be satisfied if an abrupt
change in the cosmological expansion across the spatial hypersurface
is due to a stress tensor that has a finite effect on the expansion
even in the limit of an infinitely short transition. 
We impose causality by requiring that the surface stress tensor is
also a local function of the moduli.
Related approaches have been adopted by a number of authors
\cite{DV,Nitti:2005ym,Bozza:2005qg,Chu:2006wc}. 
We use the junction conditions due to Israel \cite{Israel}
to obtain a set of equations that relate perturbation variables on
either side of the transition, in particular to consider the behaviour
across the transition of gauge-invariant quantities such as the
longitudinal gauge curvature perturbation, $\Psi$, the curvature
perturbation on uniform-density hypersurfaces, $\zeta$, and the comoving
curvature perturbation, ${\cal R}$. As a particular example we consider
the case of a single scalar field cosmology (i.e. where the moduli are
the local values of the field $\varphi$ and its velocity
$\dot\varphi$). By identifying 
the intrinsic entropy perturbation of the
single field (or isocurvature perturbation), $S_\varphi$, we find that we
require a non-zero entropy perturbation, for
the surface stress tensor to be perturbed at the transition. In other
words, we see that for the case of purely adiabatic incoming
perturbations, there will be no perturbation in the matching stress
tensor and it remains uniform over the transition surface. Armed with
this we proceed to show how the incoming perturbations on a matching
surface can be matched to the outgoing perturbations after the
transition (assuming a fluid-dominated cosmology with fixed equation
of state), relating the gauge invariant perturbations ${\cal R}$,
$\Psi$ and the intrinsic entropy perturbation $S_\varphi$. As a
result, amongst other things, we are able to understand how for the
case of purely adiabatic field perturbations the comoving curvature
perturbation remains constant across the transition surface
independent of what happens to the background expansion rate. On the
other hand, we are able to see that $\Psi$ does in general change
across the transition surface unless certain specific conditions are
satisfied.

The layout of the paper is as follows. In Section~\ref{matching} we
introduce the idea of the local matching conditions involving matching
across a spatial hypersurface (see also \cite{Nitti:2005ym}). This is
followed in section~\ref{pertns} by the introduction of the scalar
metric perturbations and the gauge invariant perturbations $\Psi$,
${\cal R}$ and $\zeta$. The junction conditions and resulting
perturbed equations are presented in section~\ref{junctions}. In
section~\ref{scalar} we apply our general analysis to the specific
case of a single scalar field cosmology, introducing the important
role of the intrinsic entropy perturbation and demonstrating how the
matching conditions depend crucially on the presence of that term. We
conclude in section~\ref{conc}.

\section{Local matching condition}
\label{matching}

Our aim is to discuss the evolution of cosmological perturbations
through a sudden transition in the cosmological evolution.
We will work in the idealised limit in which the transition is
described by matching across a spatial hypersurface, i.e., an
infinitesimally short transition. We expect this to be sufficient
to describe the evolution of perturbations on scales large with
respect to the actual duration of any transition.

We assume that the transition occurs when some function of local
observables, such as a scalar field value or the local energy density,
reaches a critical value which we describe through the
locally defined scalar function $\Sigma(\chi_I)$. The $N$ ``moduli'' $\chi_I$
schematically represent any local scalar observables. For example in the case
of a single scalar field we might consider
$\Sigma(\varphi,\dot\varphi)$ where $\dot\varphi$ is the local proper
time derivative.
We can then define the matching surface as
\begin{equation}
\Sigma(\chi_I) = 0 \,.
\end{equation}

In a spatially homogeneous FRW spacetime the matching surface must
coincide with a fixed conformal time $\eta$. In a perturbed spacetime
this need no longer be the case in general. Any 4-scalar, which is
spatially homogeneous to zeroth-order, transforms under a first-order
gauge shift as
\begin{eqnarray}
\eta &\to& \eta + \xi^0 \,,\\
\delta\Sigma &\to& \delta\Sigma -\Sigma' \xi^0 \,.
\end{eqnarray}
where $\Sigma'$ (which we require to be non-zero in the
neighbourhood of the transition) denotes the zeroth-order
derivative with respect to conformal time of $\Sigma(\chi_I)$. In
particular we can choose to work in a temporal gauge in which
$\delta\Sigma=0$, i.e., the matching surface coincides with a
constant-$\eta$ hypersurface in the perturbed spacetime.

In practice we will only be interested in the function $\Sigma$ in
the neighbourhood of the transition. All the information we need
to describe the homogeneous background transition are the values
of the moduli at $\Sigma=0$, which we will write as the parameters
$\chi_I|_\Sigma$.

To describe first-order perturbations we will also need to know
the values of the first-derivatives of $\Sigma$ with respect to
each modulus at the transition:
\begin{equation}
\Sigma_I \equiv  \left( \frac{\partial\Sigma}{\partial\chi_I}
\right)_{\Sigma=0} \,.
\end{equation}
Thus
for a study of the first-order perturbations it will
be sufficient to write $\Sigma$ in the neighbourhood of the transition
as a linear function of the moduli
\begin{equation}
\Sigma (\chi_I) = \sum_I \Sigma_I  (\chi_I - \chi_I|_\Sigma) \,.
\end{equation}
Only the study of higher-order perturbations would require knowledge of the
higher-order derivatives at $\Sigma=0$.

We will be primarily interested in cases where there is an abrupt
change in the expansion rate at the matching surface, motivated by
attempts to match cosmological perturbations through a ``bounce''
from contraction to expansion. But our results may also be applied
to other cosmological transitions such as reheating at the end of
inflation.
Any sudden change in the cosmological expansion across a spatial
hypersurface must be due to a ``singular'' stress tensor $S_{ij}$
(singular in the sense that it still has a finite effect on the
expansion in the limit of an infinitely short transition).
We can split this source term into a trace and tracefree part
\begin{equation}
\label{Sij} S_{ij} \equiv P_s q_{ij} +
a\left(\partial_i\partial_j-\frac13 q_{ij}\partial^2\right)\Pi_s
\,.
\end{equation}
The assumption that we are dealing with small perturbations about
an FRW geometry requires the stress tensor to be isotropic to
zeroth-order in the perturbations, i.e., the anisotropic stress
potential $\Pi_s$ is at most first order.

A cosmological bounce requires a violation of some of the usual
energy conditions. In particular a bounce at finite value of the
scale factor in a flat FRW universe requires violation of the null
energy condition, $\rho+P\geq0$ \cite{Brustein}.
We will restrict any such pathologies to the stress tensor
$S_{ij}$ that is localised on the matching surface. There is a
close analogy with recent studies of brane world geometries where
matter sources are assumed to be localised on a lower dimensional
spacetime (or `brane') in a higher dimensional `bulk'. The
difference in our case is that the matching surface is not
timelike, but spacelike. Matter is localised on a brane because it
is supposed to represent open string modes which at low energies
are tied to the brane. In our setting the localised stress on the
spacelike brane can include high-energy excitations that are
frozen out at low energies away from the matching surface.

The appearance of a localised source on a spacelike surface may appear
rather odd if it looks like the spontaneous appearance of some
energy-momentum out of nothing, but this need not be so. It can
approximate some actual time-evolution of the energy-momentum tensor
in the idealised limit where the timescale for this evolution is small
- infinitesimally so - with respect to all other time-scales.

Durrer and Vernizzi \cite{DV} have emphasized that if the
spacelike matching surface itself has an inhomogeneous stress
tensor then these inhomogeneities could in principle be imprinted
on the outgoing cosmology. While this is true it
could amount to simply writing on a pattern of inhomogeneities
unless this inhomogeneity can be related to pre-existing
inhomogeneities in the incoming cosmology. As a result we will
consider only models in which the localised stress tensor is a
function of local moduli.
This amounts to assuming that $P_s$ and $\Pi_s$ are scalar
functions of the $\chi_I$. In the background homogeneous solution
we only need to know the one parameter
\begin{equation}
P_s|_\Sigma \equiv P_s(\chi_I|_\Sigma) \,,
\end{equation}
as the assumption of isotropy requires
$\Pi(\chi_I|_\Sigma)=0$, while to first-order we also need
to know the $2N$ parameters
\begin{eqnarray}
P_{sI} \equiv \left( \frac{\partial P_s}{\partial \chi_I}
\right)_{\Sigma=0} \,,\\
\Pi_{sI} \equiv \left( \frac{\partial
    \Pi_s}{\partial \chi_I} \right)_{\Sigma=0} \,.
\end{eqnarray}

\section{Scalar metric perturbations}
\label{pertns}

We will write the linearly perturbed 4D metric, using the notation
of Ref.~\cite{Mukhanov:1990me}, as
\begin{equation}
ds^2 = a^2 \left[ -(1+2\phi)d\eta^2 + 2B_{,i}dx^i d\eta +
  \left\{(1-2\psi)\delta_{ij} + 2E_{,ij}  \right\} dx^i dx^j \right]
\,.
\label{ds2}
\end{equation}
where commas denote partial derivatives with respect to spatial coordinates.

The induced metric on any constant-$\eta$ hypersurface is simply the
spatial part of the metric in this gauge
\begin{equation}
\label{qij}
q_{ij} = a^2 \left[ (1-2\psi)\delta_{ij} + 2E_{,ij}  \right] \,.
\end{equation}

The extrinsic curvature of the hypersurfaces, $K_{ij}$,
may be split into a trace and tracefree part
\begin{equation}
\label{Kij}
K_{ij} \equiv \frac13 \theta q_{ij} +
a\left(\partial_i\partial_j-\frac13 q_{ij}\partial^2\right)\sigma
\,,
\end{equation}
corresponding to the expansion and shear of the orthogonal vector
field. (The vorticity vanishes for a hypersurface orthogonal vector
field.)
For the metric (\ref{ds2}) we find \cite{KS,Malik}
\begin{eqnarray}
a\theta &=& 3h(1-\phi) - 3\psi' + \partial^2\sigma \,,\\
\sigma &=& E'-B \,,
\end{eqnarray}
where $h=a'/a$ is the conformal Hubble rate and a prime denotes
derivatives with respect to conformal time.

The background expansion rate is given by the Friedmann equation
\begin{equation}
 \label{friedmann} 3h^2 = \kappa^2 a^2 \rho \,.
\end{equation}
For first-order perturbations about an FRW spacetime the Einstein
constraint equations yield
\begin{eqnarray}
3h(\psi'+h\phi) - \partial^2 (\psi+h\sigma)
 = - ah\delta\theta -\partial^2\psi
 &=& -\frac{\kappa^2}{2}a^2\delta\rho \,,
\label{pertfriedmann}\\
\psi'+h\phi = -\frac{a}{3}\delta\theta +
\frac{1}{3}\partial^2\sigma
 &=& -\frac{\kappa^2}{2}a^2 \delta q  \label{pertmtm} \,,
\end{eqnarray}
where $ \delta q$ is the momentum scalar potential.

The scalar metric perturbations $\psi$, $\phi$ and $\sigma$ are
invariant under spatial coordinate transformations $x^i\to
x^i+\partial^i\xi$, but do transform under temporal gauge
transformations:
\begin{eqnarray}
\eta &\to& \eta + \xi^0 \,,\\
\psi &\to& \psi + h\xi^0 \,,\\
\phi &\to& \phi - h\xi^0 - \xi^{0\prime} \,,\\
\sigma &\to& \sigma - \xi^0 \,.
\end{eqnarray}

We will be interested in the behaviour across the transition of
gauge-invariant quantities such as the longitudinal gauge
curvature perturbation, also known as the Newtonian or Bardeen
potential,
\begin{equation}
\label{defPsi}
\Psi \equiv \psi + h\sigma \,,
\end{equation}
the curvature perturbation on uniform-density hypersurfaces
\begin{equation}
\label{defzeta} -\zeta \equiv \psi + \frac{h}{\rho'}\delta\rho \,,
\end{equation}
and the comoving curvature perturbation
\begin{equation}
\label{defR} {\cal R} \equiv \psi - \frac{h}{\rho+P}\delta q \,,
\end{equation}
where the momentum scalar potential for a scalar field is given by
\begin{equation}
\delta q = -\frac{\varphi'\delta\varphi}{a^2} \,.
\end{equation}
It is possible to express $\zeta$ and ${\cal R}$ directly in terms
of the scalar metric perturbations using the Einstein constraint
equations (\ref{pertfriedmann}) and (\ref{pertmtm}), respectively,
to eliminate the density and momentum. This gives
\begin{eqnarray}
 \label{metriczeta}
- \zeta = \psi - \frac{2}{9(1+w)h^2}
 \left( ah\delta\theta + \partial^2 \psi \right) \,,
\\
{\cal R} = \psi - \frac{2}{9(1+w)h^2} \left( ah\delta\theta -
h\partial^2\sigma \right) \,.
 \label{metricR}
\end{eqnarray}
where we define the barotropic equation of state to be $w\equiv
P/\rho$ and make use of the continuity equation for the background fluid 
\begin{equation}
\label{fluid-eqn} \rho' + 3h(1+w)\rho=0\,.
\end{equation}
The uniform-density and comoving curvature perturbations differ by
an amount proportional to the comoving density perturbation
\begin{equation}
\label{drhoc}
 \delta\rho_c \equiv \delta\rho - 3h\delta q
 = - \frac{\rho'}{h} \left( \zeta + {\cal R} \right)
\end{equation}
which can be related to the Newtonian potential by the Einstein
constraint equations as
\begin{equation}
\label{drhoc-con}
\frac{\kappa^2}{2} a^2 \delta\rho_c = \partial^2 \Psi \,,
\end{equation}
and thus vanishes on large scales if $\Psi$ remains finite.

\section{Junction conditions}
\label{junctions}

We will choose coordinates on either side of the matching surface such
that the matching surface occurs on a constant-$\eta$ hypersurface,
i.e., $\eta=\eta_-$ on one side and $\eta=\eta_+$ on the other. This
fixes the temporal gauge at the collision on either side of the
matching. Of course, from the perturbation in this specific gauge we
can construct gauge-invariant combinations such as $\Phi$ and
$\zeta$.

Following the treatment of matching across infinitesimal shells in
general relativity \cite{Israel} we will require continuity of the first
fundamental form on the matching surface (the induced metric):
\begin{equation}
\label{match1}
\left[ q_{ij} \right]^+_- = 0 \,.
\end{equation}
Note that the extrinsic curvature is the Lie derivative of the induced
metric normal to the constant-$\eta$ hypersurface. The matching
condition (\ref{match1}) follows from assuming that the extrinsic
curvature remains finite in the limit that the thickness of the
matching surface (the duration of the transition) tends to zero. This
is gauranteed in Einstein gravity if the usual energy conditions are
obeyed \cite{Marolf}, but this will be an additional assumption if we allow
the energy conditions to be violated in the transition.
If, for example, geometry itself is only an emergent
phenomena, as sometimes claimed by string theorists, our classical
matching conditions may not be sufficient to model a truly stringy
transition. In that case we would require string theory to supply
an alternative prescription for matching fundamental degrees of
freedom on one side of the transition to the other.

The jump in the extrinsic curvature is due to the localised source
(\ref{Sij})

(see also \cite{DV})
\begin{equation}
\left[ K_{ij}  \right]^+_- = \kappa^2 \left( S_{ij} - \frac32 q_{ij}S
\right) \,.
\label{match2}
\end{equation}

Although we will consider only linearised perturbations about an FRW metric to
describe the evolution approaching or leaving the matching surface, we
could in principle describe non-linear evolution through the bounce.
The junction conditions (\ref{match1}) and~(\ref{match2}) are obtained
from the full non-linear equations of general relativity across an
interval $-\epsilon<\Delta\eta<+\epsilon$ in the ``thin-shell''
(sudden transition) limit where $\epsilon\to0$ and the extrinisic
curvature remains finite, but where the integral of the energy
momentum tensor remains finite.
This is a singular hypersurface as the Ricci tensor must diverge in
this limit in order for the integral, $S_{ij}$, to remain finite.

Equations (\ref{qij}) and (\ref{match1}) require, to zeroth-order,
continuity of the scale factor
\begin{equation}
[ a ]^+_- = 0 \,,
\end{equation}
and, to first-order in the perturbations,
\begin{equation}
[-\psi\delta_{ij} + E_{,ij} ]^+_- = 0 \,.
\end{equation}
We can split this into a trace and tracefree part which requires that
$\psi$ and $E$ are both continuous across the transition separately.  In
practice $E$ is dependent on the spatial gauge, so the physical
matching condition is that $\psi$, which describes the perturbation in
the intrinsic curvature of the spatial hypersurface, is continuous
across the surface:
\begin{equation}
\label{jumppsi}
 [\psi]^+_-= 0 \,.
\end{equation}

Equations (\ref{Kij}), (\ref{match2}) and (\ref{Sij}) require, to
zeroth-order, that the discontinuity in the Hubble expansion is due to
the background energy-momentum source
\begin{equation}
\label{jumpH}
\left[ \frac{h}{a} \right]^+_- = -\frac{\kappa^2}{2} P_{s0} \,,
\end{equation}
and, to first-order,
\begin{eqnarray}
\label{jumpdH} \left[ \frac13 \delta\theta \right]_-^+ = \left[
-\frac{1}{a}(\psi'+h\phi) +  \frac{1}{3a} \partial^2\sigma
\right]^+_- &=& -\frac{\kappa^2}{2} \delta P_s \,,\\
 \label{jumpsigma}
\left[ \sigma \right]^+_- &=&  \kappa^2 \delta\Pi_s \,,
\end{eqnarray}
where we have equated the trace and tracefree parts. Note that for
any energy-momentum tensor describing (one or many) scalar fields we have
no anistropic stress at first-order and the shear $\sigma$ must
therefore be continuous.

In order to interpret the junction condition for the trace part of the
extrinsic curvature it is useful to remember that this is related
to the energy density via the Einstein constraint equations. Using
the constraint equations (\ref{friedmann})
and~(\ref{pertfriedmann}) we can re-interpret the junction
conditions~(\ref{jumpH}) and~(\ref{jumpdH}) as describing a sudden
change in the energy density across the matching surface.

\section{Single scalar field cosmology}
\label{scalar}

In order to make this formalism more concrete we will consider the
simplest case of a single scalar field. If we demand that the
transition is triggered by a local physical quantity, this is
equivalent to demanding that $\Sigma$ is a function of the local
values of $\varphi$ and $\dot\varphi$, where to first-order we have
\begin{eqnarray}
\varphi &=& \varphi_0 + \delta\varphi \,,\\
\dot\varphi &=& \frac1a \left( (1-\phi) \varphi_0' + \delta\varphi' \right) \,.
\end{eqnarray}
Note that the perturbation of the local proper time derivative of the
field includes a term due to the perturbation of the lapse function,
$\phi$.

The condition $\Sigma_-=0$ at the bounce then imposes the constraints
\begin{eqnarray}
\label{dSigvarphi}
\Sigma_1 \delta\varphi_- + \frac1a \Sigma_2
(\delta\varphi_-'-\varphi_{0-}'\phi_-) &=& 0 \,,
\end{eqnarray}
so there is a linear relation between the first-order perturbations of
the field and its derivative at the transition, fixed by
the ratio $\Sigma_1/\Sigma_2$.

Similarly we assume that the surface stress is a function of the local
field and its proper time derivative so that
\begin{eqnarray}
\label{Psvarphi}
P_s(\varphi,\dot\varphi) = P_{s0} + P_{s1}\delta\varphi_- + \frac1a P_{s2} (\delta\varphi_-'-\varphi_0'\phi_-) \,,\\
\label{Pisvarphi} \delta\Pi_s(\varphi,\dot\varphi) =
\Pi_{s1}\delta\varphi_- + \frac1a \Pi_{s2}
(\delta\varphi_-'-\varphi_0'\phi_-) \,,
\end{eqnarray}

Using the constraint (\ref{dSigvarphi}) we can write
\begin{eqnarray}
\label{dPsvarphi}
\delta P_s &=& \left( P_{s1} - \frac{\Sigma_1}{\Sigma_2} P_{s2} \right) \delta\varphi_- \,,
\\
\label{dPisvarphi}
\delta\Pi_s &=& \left( \Pi_{s1} -  \frac{\Sigma_1}{\Sigma_2} \Pi_{s2} \right) \delta\varphi_- \,.
\end{eqnarray}

\subsection{Non-adiabatic perturbations}

A local rotation of the perturbations in phase-space is a useful
technique to understand the evolution of large-scale perturbations in
cosmology by identifying adiabatic and entropy perturbations
\cite{Gordon}. We define adiabatic perturbations to be perturbations
along the zeroth order (background) trajectory in phase space for all
variables. Thus to first order we have
\begin{equation}
\frac{\delta x}{\dot{x}} = \frac{\delta y}{\dot{y}} \ \forall\  x,y \,.
\end{equation}
Any perturbation orthogonal to the background trajectory represents a
relative entropy (or isocurvature) perturbation
\begin{equation}
S_{xy} = \frac{h}{a}
  \left( \frac{\delta x}{\dot{x}} - \frac{\delta y}{\dot{y}} \right) \,,
\end{equation}
where we include the local Hubble rate $h/a$ to make $S_{xy}$ dimensionless.

For a single scalar field described by the two-dimensional phase-space
$(\varphi,\dot{\varphi})$ we can define the intrinsic entropy
perturbation of the single field
\begin{equation}
\label{Svarphi}
S_\varphi = \frac{h}{a}
 \left(
\frac{\dot{\delta\varphi}}{\ddot\varphi_0} - \frac{\delta\varphi}{\dot\varphi_0} \right)
 \,,
\end{equation}
where remember that a dot denotes differentiation with respect to
proper time. Thus we can re-write this in terms of conformal time
derivatives as
\begin{equation}
S_\varphi = h \left( \frac{\delta\varphi'-\varphi'\phi}{\varphi''-h\varphi'} - \frac{\delta\varphi}{\varphi'} \right) \,.
\end{equation}

The comoving density perturbation for a single scalar field is
\begin{equation}
\delta\rho_c = \frac{1}{a^2} \left( \varphi'(\delta\varphi' -\varphi'\phi)-(\varphi''-h\varphi')\delta\varphi \right) \,,
\end{equation}
and hence is related to the relative entropy (\ref{Svarphi})
\begin{equation}
\delta\rho_c = \frac{\varphi'(\varphi''-h\varphi')}{a^2h} S_\varphi \,.
\end{equation}
This is in turn related to the Newtonian potential via the
Einstein constraint equation (\ref{drhoc-con}), and hence for a
single scalar field we have
\begin{eqnarray}
 \label{SPsi}
S_\varphi &=&
\frac{2h}{\kappa^2\varphi'(\varphi''-h\varphi')}
\partial^2 \Psi \,,
 \nonumber\\
&=&-\frac{4}{9h^2(1+w)(1+c_s^2)}
\partial^2 \Psi \,.
\end{eqnarray}
where we have written $w=P/\rho$ and $c_s^2=\dot{P}/\dot\rho$ for
the scalar field energy and pressure.

We can re-write Eq.~(\ref{dSigvarphi}) as a constraint relating the
field fluctuations on the transition surface to the instrinsic entropy
fluctuation
\begin{equation}
\left(\Sigma_1 + \frac{\varphi''-h\varphi'}{a\varphi'} \Sigma_2
\right) \delta\varphi_- = - \Sigma_2 \left(
  \frac{\varphi''-h\varphi'}{ah} \right) S_\varphi \,.
\end{equation}
where we have used Eq.~(\ref{Svarphi}) to eliminate
$\delta\varphi'-\varphi_0'\phi$.

We will write this as
\begin{equation}
\label{dvarphiin}
\delta\varphi_- = - \s_\Sigma \frac{\varphi'}{h} S_{\varphi-} \,,
\end{equation}
where
\begin{equation}
\s_\Sigma  \equiv \frac{\Sigma_2 \left(\varphi''-h\varphi'\right)}
{\Sigma_1 a\varphi' + \Sigma_2 \left(\varphi''-h\varphi'\right)} \,.
\end{equation}
The dimensionless ratio $\s_\Sigma$ defines the rate of change of
$\Sigma$ due to the change of $\varphi$ versus that due to the change
$\dot\varphi$.
We see that in the absence of intrinsic entropy fluctuations
($S_{\varphi-}=0$) there can be no field perturbations at the matching
surface, $\delta\varphi_-=0$. \footnote{Note that we require
  $\dot\Sigma=\Sigma_1\dot\varphi+\Sigma_2\ddot\varphi\neq0$ for
  transition to occur.}

{}From Eqs.~(\ref{dvarphiin}), (\ref{dPsvarphi}) and
(\ref{dPisvarphi}) we have
\begin{eqnarray}
\label{dPsS}
\delta P_s &=& - \left( \s_P - \s_\Sigma \right)
 \frac{{P}_s'}{h_-} S_{\varphi-} \,,
\\
\label{dPisS}
\delta\Pi_s &=& - \left( \s_\Pi - \s_\Sigma \right)
 \frac{\Pi_s'}{h_-} S_{\varphi-} \,,
\end{eqnarray}
where we have defined
\begin{eqnarray}
\s_P &\equiv& \frac{P_{s2}\left(\varphi''-h\varphi'\right)}
{P_{s1} a\varphi' + P_{s2} \left(\varphi''-h\varphi'\right)} \,,
\\
\s_\Pi &\equiv& \frac{\Pi_{s2}\left(\varphi''-h\varphi'\right)}
{\Pi_{s1} a\varphi' + \Pi_{s2} \left(\varphi''-h\varphi'\right)} \,.
\end{eqnarray}
Thus we see from Eqs.~(\ref{dPsvarphi}) and (\ref{dPisvarphi})
that we require a non-zero intrinsic entropy perturbation for the
surface stress tensor to be perturbed at the transition.
For purely adiabatic incoming perturbations the matching surface
stress tensor is necessarily unperturbed.

\subsection{Incoming perturbations on matching surface}

We can now express the metric perturbations $\psi_-$,
$\delta\theta_-$ and $\sigma_-$ on the matching surface in terms
of the incoming field perturbations $\delta\varphi_-$ and the
gauge-invariant curvature perturbations ${\cal R}_-$ and $\Psi_-$.

{}From the definition of the comoving curvature perturbation, ${\cal
  R}$ in Eq.(\ref{defR}), we can write the curvature perturbation on
the matching surface as
\begin{equation}
\label{dpsi-}
\psi_- = {\cal R}_- - \frac{h_-}{\varphi'}\delta\varphi_- \,.
\end{equation}
On the other hand, the shear perturbation on the matching surface can
be written, using Eq.~(\ref{defPsi}), in terms of $\psi_-$ and the
Newtonian potential
\begin{equation}
 \sigma_- = \frac{1}{h_-} \left( \Psi_- - \psi_-
\right) \,,
\end{equation}
and hence
\begin{equation}
 \label{sig-}
\sigma_- = \frac1h \left( \Psi_- - {\cal R}_- +
\frac{h_-}{\varphi'} \delta\varphi_- \right) \,.
\end{equation}

Finally, the perturbed expansion $\delta\theta$ is related to the
field perturbation via the momentum constraint~(\ref{pertmtm}).
\begin{equation}
 a\delta\theta_- =
 - \frac{3\kappa^2}{2}\varphi'\delta\varphi_- + \partial^2\sigma_-
 \,,
\end{equation}
and hence
\begin{equation}
 \label{dtheta-}
a\delta\theta_- = - \frac{3\kappa^2}{2}\varphi'\delta\varphi_- +
 \frac{1}{h_-} \partial^2 \left( \Psi_- - {\cal R}_- +
 \frac{h_-}{\varphi'} \delta\varphi_- \right) \,.
\end{equation}

Equations~(\ref{dpsi-}), (\ref{sig-}) and~(\ref{dtheta-}) give the
geometrical perturbations on the incoming side of the matching surface
in terms of the field fluctuations $\delta\varphi_-$ and the
gauge-invariant metric perturbations $\R_-$ and $\Psi_-$. However we have
seen that the field perturbations on the matching surface
$\delta\varphi_-$ can be related to the gauge-invariant entropy
perturbations $S_-$ for a single fluid through Eq.~(\ref{dvarphiin}).
Thus we have
\begin{eqnarray}
\label{inpsi}
 \psi_- &=& \R_- +\s_\Sigma S_{\varphi-} \,,\\
\label{intheta}
 a \delta\theta_- &=&
 \frac{9(1+w_-)h_-}{2}
 \s_\Sigma S_{\varphi-}
  +\frac{1}{h_-}\partial^2 \left(\Psi_- - \R_- - \s_\Sigma
 S_{\varphi-} \right)
 \,,\\
\label{insigma}
 \sigma_- &=& \frac{1}{h_-} \left( \Psi_- - \R_- - \s_\Sigma
 S_{\varphi-} \right) \,.
\end{eqnarray}

We note that in the absence of entropy perturbations we have
\begin{eqnarray}
\psi_- &=& {\cal R}_- \,, \\
a\delta\theta_- &=& \frac{1}{h_-} \partial^2 \left( \Psi_- - {\cal R}_- \right) \,,\\
\sigma_- &=& \frac{1}{h_-} \left( \Psi_- - {\cal R}_- \right) \,.
\end{eqnarray}
Although we have written the three metric perturbations on the
matching surface $\psi_-$, $\delta\theta_-$ and $\sigma_-$ in
terms of the three gauge-invariant variable ${\cal R}_-$, $\Psi_-$
and $S_{\varphi-}$, two of the gauge-invariant variables, $\Psi_-$
and $S_{\varphi-}$, are related by the constraint
Eq.~(\ref{SPsi}). This requires the entropy perturbation to be
much smaller than $\Psi$ on large (typically super-Hubble) scales.

\subsection{Outgoing perturbations}

We will now reconstruct the outgoing perturbations after the
transition assuming a fluid dominated cosmology with fixed
equation of state $w_+\equiv(P/\rho)_+$.

The outgoing comoving curvature perturbation is written using
Eq.~(\ref{metricR}) as
\begin{equation}
{\cal R}_+ = \psi_+ + \frac{2}{9(1+w_+)h_+} \left(
-a\delta\theta_+ + \partial^2 \sigma_+ \right) \,.
\end{equation}
and thus using the junction conditions (\ref{jumppsi}),
(\ref{jumpdH}) and~(\ref{jumpsigma}) we have
\begin{equation}
{\cal R}_+ = \psi_- + \frac{2}{9(1+w_+)h_+} \left(
-a\left[\delta\theta_- - \frac32 \kappa^2\delta P_s\right] + \partial^2
\left[ \sigma_-+\kappa^2\delta\Pi_s \right] \right) \,.
\end{equation}
We can write this in terms of the incoming gauge-invariant
perturbations with a single scalar field, using Eqs.~(\ref{dPsS}),
(\ref{dPisS}), (\ref{inpsi}) and~(\ref{intheta}), as
\begin{eqnarray}
{\cal R}_+ &=& {\cal R}_- + \left[ \left(
1-\frac{(1+w_-)h_-}{(1+w_+)h_+} \right) \s_\Sigma \right.
\nonumber\\
 && \left. - \frac{2\kappa^2}{9(1+w_+)h_+} \left(
\frac{3a}{2}\frac{P_s'}{h_-}(\s_P-\s_\Sigma) +
\frac{\Pi_s'}{h_-} (\s_\Pi - \s_\Sigma )
\partial^2 \right) \right] S_{\varphi-} ,.
\end{eqnarray}

The outgoing gauge-invariant perturbation $\Psi_+$ is given from
Eq.~(\ref{defPsi}) as
\begin{equation}
\Psi_+ \equiv \psi_+ + h_+\sigma_+ \,.
\end{equation}
The junction conditions (\ref{jumppsi}) and~(\ref{jumpsigma}) then
give
\begin{equation}
\Psi_+ = \psi_- + h_+\left(\sigma_- + \kappa^2\delta\Pi_s \right)
 \,.
\end{equation}
Again we can write this in terms of the incoming gauge-invariant
perturbations with a single scalar field, using
Eqs.~(\ref{dPisS}), (\ref{inpsi}) and~(\ref{insigma}), as
\begin{equation}
\Psi_+ = \frac{h_+}{h_-}\Psi_- - \frac{h_+-h_-}{h_-} \left( {\cal
R}_-
  + \s_\Sigma S_{\varphi-} \right)
 - \frac{h_+}{h_-} \kappa^2 \Pi_s' (\s_\Pi-\s_\Sigma)
 S_{\varphi-}
   \,.
\end{equation}

For the special case of purely adiabatic field perturbations we
have the simple result
\begin{eqnarray}
\label{simpleR}
 {\cal R}_+ &=& {\cal R}_- \,,\\
\label{simplePsi}
 \Psi_+ &=& \frac{h_+}{h_-}\Psi_- + \left(1-\frac{h_+}{h_-}\right) {\cal R}_-
  \,.
\end{eqnarray}
Thus the comoving curvature perturbation is constant for adiabatic
perturbations across an arbitrary transition surface,
$\Sigma(\varphi,\dot\varphi)=0$, even if there is an abrupt change in
the background expansion rate $h$, or if the localised stress-tensor
on the transition surface is an arbitrary function of the scalar field
or its proper time derivative. In contrast, the Newtonian potential
does change across the transition, unless there is no change in the
background expansion rate, i.e., $h_+=h_-$, or the special case of
vanishing comoving shear, in which case $\Psi={\cal R}$.

The adiabatic limit is a very restrictive case, but it is one that
is a good approximation in cases where there is a unique attractor
in phase-space and the scalar field perturbations are in a
squeezed state, as is the case during slow-roll inflation, or an
ekpyrotic-type collapse.

In practice the intrinsic entropy perturbation of the scalar field
is related to the divergence of the Newtonian potential through
Eq.~(\ref{SPsi}). Thus we can write the general transfer matrix
through a sudden transition as
\begin{equation}
\left( \begin{array}{c} {\cal R}_+ \\ \Psi_+ \end{array} \right)
 =
 \left( \begin{array}{cc} T_{RR} & T_{R\Psi} \\ T_{\Psi R} & T_{\Psi\Psi} \end{array} \right)
 \left( \begin{array}{c} {\cal R}_- \\ \Psi_- \end{array} \right)
 \,.
\end{equation}
where we can split the transfer matrix into terms of order $k^0$,
$k^2$ and $k^4$:
\begin{equation}
 \label{Tijn}
T_{ij} = \sum_{n=0,1,2} k^{2n} T_{ij}^{(2n)} \,.
\end{equation}
On large scales we expect the outgoing perturbations to be determined
by the $k^0$-coefficients:
\begin{eqnarray}
 T_{RR}^{(0)} &=& 1 \,,\\
 T_{R\Psi}^{(0)} &=& 0 \,,\\
 T_{\Psi R}^{(0)} &=& \frac{h_--h_+}{h_-} \,,\\
 T_{\Psi\Psi}^{(0)} &=& \frac{h_+}{h_-}  \,.
\end{eqnarray}
In particular we see that, even if we allow the surface
energy-momentum tensor to be an arbitrary function of the local field
and its proper time derivative, the comoving curvature perturbation is
conserved in this large-scale limit.

On finite scales the full transfer matrix includes the additional
non-zero coefficients:
\begin{eqnarray}
T_{R\Psi}^{(2)} &=& \frac{4}{9h_-^2(1+w_-)(1+c_{s-}^2)} \left[
\left(
1-\frac{(1+w_-)h_-}{(1+w_+)h_+} \right) \s_\Sigma -
\frac{\kappa^2aP_s'}{3(1+w_+)h_+h_-} (\s_P-\s_\Sigma) \right]  \,,
\\
T_{R\Psi}^{(4)} &=& \frac{4}{9h_-^2(1+w_-)(1+c_{s-}^2)} \left[
 \frac{2\kappa^2}{9(1+w_+)h_+} \left(
\frac{\Pi_s'}{h_-} (\s_\Pi - \s_\Sigma )
 \right) \right] \,,\\
T_{\Psi\Psi}^{(2)} &=& - \frac{4}{9h_-^2(1+w_-)(1+c_{s-}^2)}
 \left[ \frac{h_+-h_-}{h_-} \s_\Sigma
 + \frac{h_+}{h_-} \kappa^2 \Pi_s' (\s_\Pi-\s_\Sigma) \right] \,.
\end{eqnarray}
All the other coefficients in Eq.~(\ref{Tijn}) are zero.
If all the functions $\Sigma$, $P_s$ and $\Pi_s$ are functions only of
the local value of the scalar field, $\varphi$, and not its time
derivative, then $\s_\Sigma$, $\s_P$ and $\s_\Pi$ all vanish and only
the $k^0$ coefficients $T_{ij}^{(0)}$ are non-zero.

\section{conclusions}
\label{conc}

In this paper we have presented the general cosmological matching
conditions for linear scalar metric perturbations at a sudden
transition across a spacelike hypersurface, allowing for a perturbed
surface stress tensor leading to a finite change in the extrinsic
curvature. This describes the behaviour of perturbations on scales
much larger than the characteristic length associated with causal
propagation during a cosmological transition such as the end of
inflation, or a bounce at the end of a pre big bang-type model.

Durrer and Vernizzi \cite{DV} emphasized that an inhomogeneous
surface stress at the transition might lead to a change in the
comoving curvature perturbation, ${\cal R}$. This could be of great
importance in the ekpyrotic or cyclic scenarios where the comoving
curvature perturbation during the pre big bang phase does not have a
scale-invariant spectrum \cite{DHL}, even though the Newtonian
potential, $\Psi$, does. 

We studied the case of a sudden transition from a universe dominated
by a single scalar field to a radiation-dominated era, giving
expressions for the outgoing gauge-invariant perturbations $\Psi$
and ${\cal R}$ in terms of the incoming perturbations. We found that
if the matching surface (both the transition time and the surface
stress tensor) can be characterised by functions of the local scalar
field and its proper time derivatives then any change in the
comoving curvature perturbation on large scales must be due to
intrinsic entropy pertubations in the scalar field. However during
slow-roll inflation or an ekpyrotic collapse the scalar field is
driven into a squeezed state on super-Hubble scales and entropy
perturbations are heavily suppressed. In particular the intrinsic
entropy perturbation of a scalar field is proportional to the
divergence of the Newtonian potential, $\partial^2 \Psi$, and has a
steep blue spectrum even when the Newtonian potential is almost
scale invariant. 

If we simplify the analysis by considering only adiabatic
perturbations then the surface stress tensor must be unperturbed, and
we obtain 
\begin{eqnarray}
 \left[ {\cal R} \right]^+_- &=& 0 \,,\\
 \left[ \Psi \right]^+_- &=& \left( \frac{H_+-H_-}{H_-} \right) \left(
  \Psi_- - {\cal R}_- \right) \,.
\end{eqnarray}
If the background expansion is continuous, $[H]^+_-=0$, then we
 recover the standard result that both ${\cal R}$ and $\Psi$ are both
 continuous \cite{HV,DM,MS}.
A sudden transition in the Hubble expansion can lead to a sudden
jump in the Newtonian potential $\Psi$, but the comoving curvature
perturbation remains constant.
This is in accordance with the general rule that there exists a conserved
curvature perturbation for adiabatic perturbations on large scales so
long as local energy conservation holds~\cite{WMLL,LW03}.

Our results re-inforce previous claims (see, e.g., \cite{DHL}) based on
the evolution of cosmological perturbations in general relativity, 
and recent studies \cite{Creminelli,Bozza:2005qg,Chu:2006wc} which seek to go
beyond general relativity.
We emphasise that our conclusions apply to any model of a sudden
transition between two epochs described by general relativity, even if
general relativity breaks down during the transition, so long as
physics is still local and the transition is described in terms of
local moduli or matter variables.

\section*{Acknowledgements}

DW is grateful to Antonio Cardoso for his careful reading of the
manuscript and useful comments.
DW is partly supported by PPARC grants PPA/G/S/2002/00576 and PP/C502514/1.
We would like to thank the organisers of the Kavli Institute for
Theoretical Physics workshop on Superstring Cosmology, for their
hospitality during the summer of 2003 (yes, really!) when this
project began, and the organisers of the Benasque workshop on Modern
Cosmology in August 2006 when it was completed.

\end{document}